# Influence of Sequential Changes in the Crude Oil-Water Interfacial Tension on Spontaneous Imbibition in Oil-Wet Sandstone

Anupong Sukee[1], Tanakon Nunta[1], Maje Alhaji Haruna[2],
Azim Kalantariasl[3], and Suparit Tangparitkul[1,*]

1) Department of Mining and Petroleum Engineering, Faculty of Engineering, Chiang Mai University, Chiang Mai, 50200, Thailand

2) School of Chemical and Process Engineering, University of Leeds, Leeds, LS2 9JT, United Kingdom

3) Department of Petroleum Engineering, School of Chemical and Petroleum Engineering, Shiraz University, Shiraz, 71946-84471, Iran

*To whom correspondence should be addressed: Email suparit.t@cmu.ac.th Phone: +66 5394 4128 Ext. 119

**Abstract**

Crude oil-water interfacial tension in petroleum reservoir is reduced or increased due to surfactant injection or surfactant retention, respectively. Changes in the interfacial tension crucially attribute to a governing capillary pressure and hence an oil displacement in spontaneous imbibition process. While a reduction in the interfacial tension has been highlighted as one of the underlying mechanisms for enhanced oil recovery, fluctuated surfactant concentration within reservoir promptly disturbs such interfacial phenomenon. The current study therefore attempts to elucidate an influence of such changes on spontaneous imbibition by replacing surfactant concentration consecutively with two approaches: sequential decrease and sequential increase in the interfacial tension. Two fluid flow directions were examined simultaneously: co-current and counter-current flows. Dimensionless numbers were analyzed to emphasize the fluid displacement. With strongly oil-wet wettability (contact angle ≥ 123°), capillarity was a resisting element to oil displacement and therefore controlled by the oil-water interfacial tension. The sequential-reduced interfacial tension was found to weaken such resisting capillary force gradually and resulted in consecutive incremental oil production. On the contrary, the sequential-increased interfacial tension initiated the lowest interfacial tension fluid that produced an immediate large amount of oil, but did not much displace further oil. The current study also observed a greater oil recovery obtained from a sequential reduction in the interfacial tension scheme (26.9%) compared to a conventional single reduction scheme (22.4%), with both schemes attaining same interfacial tension at last. Variation in pore-filling events was believed to attribute to such discrepancy since an inertia hinderance to oil displacement developed differently. In counter-current imbibition, same characteristics of oil displacement were observed as in co-current imbibition, with less oil produced (≤ 17.6% ultimate recovery) and less sensitive to fluid changes due to negligible gravitational contribution. The results emphasized how the sequential-reduced interfacial tension exhibits a greater oil recovery by imbibition as analogy to secondary oil production by surfactant injection after water flooding, while increasing interfacial tension is likely attributed from surfactant retention could produce less oil.

**Highlights**

- Appropriate reduction in interfacial tension produces greater oil with high rate.
- Sequentially reduced interfacial tension produces more oil than a single-reduced.
- Sequential increase in interfacial tension yields less oil production with snap-off.
- Gravitational and inertia effects also influence capillary-dominated oil imbibition.

# 1 Introduction

Change in the interfacial tension ($\sigma$) between crude oil and connate water from pristinely high to relatively low value has long been understood to improve petroleum production, i.e. enhanced oil recovery (EOR), after primary and secondary recoveries terminated. Chemical additives are hence implemented with aim to reduce the $\sigma$, e.g. surfactant injection [1]–[4] Surfactants could effectively reduce $\sigma$ to a magnitude of $10^{-3}$ mN/m [5], [6], and this reduction induces trapped crude oil to be displaced from porous reservoir rock via a number of related mechanisms reported (e.g. decrease in oil-rock adhesion work, spontaneous emulsification, and wettability alteration) [2], [7]–[13].

Considering interfacial phenomena in oil reservoir controlled by capillarity, surfactant-reduced $\sigma$ could improve an oil displacement in an unfavorable oil-wet reservoir [11], [14]. Capillary pressure ($P_c$) describing such phenomena is expressed as a function of oil-water-rock interactions as [15]:

$$P_c = \frac{2\sigma cos\theta}{r} \tag{1}$$

where $\theta$ is the three-phase oil-water-rock contact angle representing porous media wettability, and $r$ the rock pore radius. With Eq. (1), wettability ($\theta$) is thus inevitably the main dominating factor since it governs a direction of capillary pressure: water-wet reservoir ($\theta < 90°$) delivers a positive capillary pressure (a driving force to oil displacement); and oil-wet reservoir ($\theta > 90°$) facilitates a negative capillary pressure that resists oil displacement.

Although $\sigma$ values are screened by wettability, reducing $\sigma$ by surfactant addition could significantly enhance oil recovery by either (i) $\sigma$ reduction in oil-wet reservoir to diminish a resisting capillary or (ii) to promote wettability alteration toward water-wet (due to Young's equation), but not both concurrently. Normally, no $\sigma$ reduction or additional surfactant is required in water-wet reservoir to ensure a driving capillary force remained [11].

Surfactant injection for $\sigma$ modification is crucial to stimulate in-situ spontaneous imbibition which plays a key role in EOR, especially for tight and fractured reservoirs (e.g. tight rocks and shale oil) [11], [16]–[21]. Spontaneous imbibition is a natural-driven phenomenon that wetting fluid displaces non-wetting fluid, controlled primarily by capillary force and additionally some others [22]–[24].

There are discrepancies reported in $\sigma$ change and its resulted oil imbition, even though $\sigma$ reduction by surfactants has been widely documented with a robust potential to improve oil recovery and regarded as one of the main EOR mechanisms [2], [25]–[28].

Reduced $\sigma$ was observed to increase oil recovery by many research groups. Gao et al. [4] investigated the TOF-1 surfactant at various concentrations in fracturing fluid and found that lower $\sigma$ produced higher oil recovery. With several surfactants investigated, Hou et al. [29] have also observed that the highest oil imbibition (< 19.3% recovery) was produced from the lowest $\sigma$ (0.226 mN/m) by using 0.2% cetrimonium bromide (CTAB) surfactant. However, some studies reported a controversial result affected by $\sigma$. Milter and Austad [30] found only 5% oil recovery with substantially low $\sigma$ (0.02 mN/m) from imbibition experiments. The authors argued that much reduced $\sigma$ by surfactant appeared to prevent water imbibing into rock sample under mixed-wet condition. While mechanisms remain unclear, these findings suggested that extremely reduced $\sigma$ is not necessarily an optimal for imbibition recovery and may even hinder an oil displacement.

Equation (1) implies contribution of high $\sigma$ to large driving capillary in water-wet rock and therefore a promising high imbibition recovery, however high $\sigma$ values also result in large oil-rock adhesion force ($F_A = \pi R \sigma \sin\theta$, where $R$ is the radius of oil-rock contact area) which opposes to oil detachment from rock surface [2]. This suggests that there is an 'optimal' $\sigma$ value or a range that delivers maximum oil recovery, not as low or as high as possible. A recent study by You et al. [31] on synergistic contribution from wettability and $\sigma$ confirmed an optimal $\sigma$ value for highest oil recovery, with wettability remained water-wet. The study found that as $\sigma$ increasing (from $2.7 \times 10^{-3}$ to 19.5 mN/m) the imbibition recovery initially increased and then decreased, with the highest oil recovery obtained at an intermediate $\sigma$ value ($9.5 \times 10^{-2}$ mN/m). Ren et al. [32] also found an optimal $\sigma$ of 0.87 mN/m contributing to maximum oil imbibition from an interfacial tension range of 0.32 – 10.82 mN/m, highlighting that the lowest $\sigma$ does not always favor an oil displacement. Alvarez et al. [33] further verified that a moderate-$\sigma$-reducing surfactant contributed to much effective wettability alteration, and thus resulted in a better oil displacement.

Although effect of $\sigma$ change by surfactant injection has been widely reported, some limited studies related to field implementation are yet to be explored. Conventionally, surfactant injection via an injecting well is aimed to sweep or 'flood' residual oil in targeted zone of reservoir toward a producing well. With lengthy travelling path, surfactant concentration is argued to fade once arrives at reservoir target due to surfactant retention (surfactant adsorption at the rock-water interface) [34], [35]. Such surfactant loss could accordingly lead to an ineffective $\sigma$ reduction (i.e. $\sigma$ increased) and thus results in EOR capability. Further studies are therefore needed to systematically examine an influence of $\sigma$ increase as contributed from decreasing surfactant concentration. Oil displacement from spontaneous imbibition process as contributed from the $\sigma$ changes is of interest in the current study. Resulted findings could deliver some insightful understandings to design a supplementary chemical flooding weather high or low $\sigma$ is required to displace further oil after one has been injected and ceased to function.

In the current study, oil recovery in spontaneous imbibition process was investigated on an effect of sequential $\sigma$ change. By means of replacing surfactant solutions consecutively, the $\sigma$ change was designed in two approaches: (i) sequential decrease and (ii) sequential increase, mimicking (i) surfactant flooding and (ii) its following water flooding or surfactant retention, respectively. Imbibition test was also examined in two fluid flow directions: co-current and counter-current. Resulted oil recoveries from each experimenting condition were elucidated with the measured oil-water-rock microscopic properties ($\sigma$ and $\theta$) and relevant dimensionless numbers, including emphasis on pore-filling events that were believed to influence oil displacement behaviors.

## 2 Materials and Experimental Methods

### 2.1 Oil phase and surfactant

Crude oil was collected from a primary-production well (FA-BT63-10) at Fang oilfields in Chiang Mai (Thailand), and its properties are shown in Table 1. The crude oil was shaken and de-gassed before use. To better dissolve wax components and minimize viscous effect, the crude oil was mixed with toluene (RCI Labscan, Thailand) at 1:1 ratio by volume and used as an oil phase throughout the experimental study (henceforth referred to as oil phase). The mixed oil has density of 856.5 kg/m$^3$ and viscosity of 3.18 mPa·s at 40 °C.

**Table 1.** Crude oil properties.

| Density at 40 °C (kg/m$^3$) | Viscosity at 40 °C (mPa·s) | Wax content (wt%) | Asphaltene content (wt%) | Total acid number (mg KOH/g) |
|---|---|---|---|---|
| 864.7 | 22.0 | 51.0 | 0.05 | 0.10 |

Surfactant used in the current study was Triton X-100 (Loba Chemie, India), a non-ionic surfactant, with molecular weight of 647 g/mol. Our previous study found that the critical micelle concentration (CMC) of Triton X-100 was 0.22 mM [34]. CMC is the critical surfactant concentration that obtains minimum $\sigma$ value. If surfactant concentration continuously increases beyond CMC, $\sigma$ effect could not be investigated since $\sigma$ values remain constant. Previous studies found a negligible Triton X-100 contribution to change the contact angle due to a reduced oil-water interfacial tension (via Young's equation [15]) [36], hence a sole contribution from a reduction in the oil-water interfacial tension can be studied specifically. Surfactant solutions were prepared with deionized water (RCI Labscan, Thailand).

## 2.2 Sandstone core samples and aging process

Four core samples (5 cm diameter and 5 cm length) were drilled from a single block of sandstone outcrop, taken from Khorat Group in Nong Khai (Thailand), to ensure that the rock properties are consistent. The rock lithology mainly consists of quartz. The rock samples were visually assessed to be homogenous in constituent minerals and pore configuration, without any laminations observed. Microscopic images of the rock surface were used to determine pore size distribution using an image processing program (ImageJ), see Fig. 1a. Two ranges of equivalent pore diameters were observed (Inset of Fig. 1a), and the average diameters are 255 ± 15 and 99.8 ± 2.3 μm, which are presumed to be pore and throat sizes of the rock samples, respectively.

Effective porosity of the core samples was determined by liquid saturation method using deionized water. The porosity of all core samples was consistent of ~10%, a fairly low porosity for clastic rock. Measured porosity indicates an unconventional reservoir characteristic, which capillary force normally controls fluid flow behavior [37].

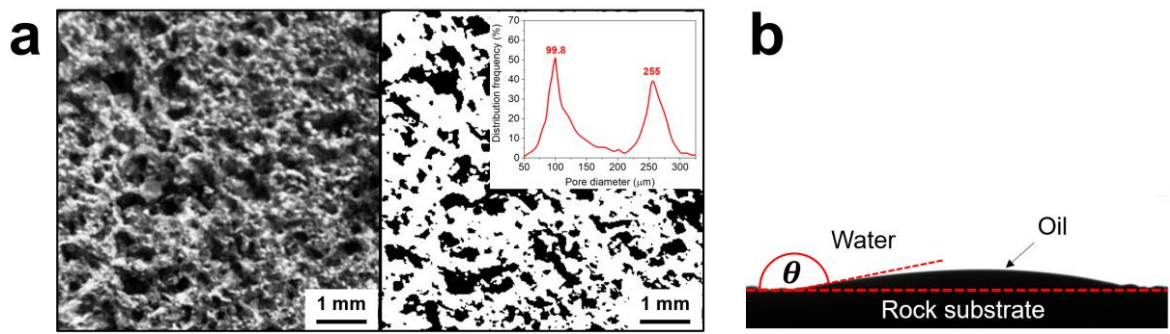

**Figure 1.** Microscopic images of the rock surface (left) and its processed binary image (right) for pore size characterization (a). Inset is the pore-size distribution curve showing two peaks for average sizes of pore and throat. Rock initial wettability (after 8-day aging with the oil phase) characterized by the oil-water-rock contact angle (b).

Prior to imbibition experiment (Section 2.4), each core sample was cleaned by saturating with deionized water and left to dry overnight in an oven. Cleaned core samples were thereafter saturated with the oil phase under vacuum condition for 20 min and then aged at 40 °C for 8 days. After saturated, an initial oil saturation in each sample was determined by means of weight difference. Owing to adsorption of crude oil components on rock surface, the aging process has established an initial wettability of rock surface to be strongly oil-wet characterized by the three-phase contact angle of 171.8° (Fig. 1b).

## 2.3 Measurements of the oil-water interfacial tension and the three-phase contact angle

The interfacial tension between the oil phase and aqueous phase ($\sigma$) was measured by a pendent drop technique using an Attension® Theta Optical Tensiometer (TF300-Basic, Biolin Scientific, Finland) equipped with a thermal cell (C217W, Biolin Scientific, Finland) and operated at 40 ± 1 °C. An oil droplet of 10 µL was dispensed at the tip of a stainless inverted needle (gauge 22) using a micro-syringe pump (C201, Biolin Scientific, Finland). The shape of the oil droplet was recorded at 3.3 fps until no detectable change was observed (~1,200 s).

The three-phase oil-water-rock contact angle ($\theta$) was measured to characterize the rock surface wettability by an Attension® Theta Optical Tensiometer (TF300-Basic, Biolin Scientific, Finland) operated at 40 ± 1 °C. Rock substrates (~3 mm thickness) were cut from the same block of sandstone outcrop, thereafter cleaned and aged with the oil phase as described previously. The aged substrate was stationed on a home-made acrylic stand placed inside a thermal cell (C217W, Biolin Scientific, Finland), with the rock substrate hang in a center of the cell window for contact angle observation. Aqueous solution was pre-heated to 40 °C and added to the thermal cell to submerge the rock substrate. An oil droplet of 5 µL was deposited underneath the rock substrate by means of an inverted needle attached to a micro-syringe, and the oil-water-rock contact angle was formed. The contact angle was recorded at 3.3 fps until the steady-state contact angle (no detectable change) was observed (~2 h). The three-phase contact angle was measured through the water phase from image analysis using the OneAttension software. The average contact angle (contact angles measured from the left and right) is reported.

For both measurements, the aqueous phase was deionized water or surfactant solutions at various concentrations. The measurements were performed in triplicate and the average values were reported.

## 2.4 Spontaneous imbibition design and experiment

### 2.4.1 Sequential changes in the oil-water interfacial tension

To investigate an influence of sequential changes in the oil-water interfacial tension ($\sigma$) on spontaneous imbibition, $\sigma$ values were controlled by surfactant concentrations of four orders of magnitude (i.e. 0, 0.002, 0.02, 0.2 mM Triton X-100), while the sequential changes in $\sigma$ were achieved by replacing surfactant solution by other concentrations in sequence.

Study on the sequential changes in $\sigma$ consisted of two approaches: sequential increase and sequential decrease in $\sigma$. It is noted that surfactant concentration is inversely proportional to the $\sigma$ value in general. To decrease $\sigma$ sequentially, surfactant fluids for imbibition experiment is changed in order of sequentially increasing concentration: 0 → 0.002 → 0.02 → 0.2 mM; and vice versa for another approach (sequential increase in $\sigma$). Consecutive order of surfactant fluids for each approach is referred to as: Fluid I, Fluid II, Fluid III, and Fluid IV, as annotated in Table 2.

**Table 2.** Sequential fluid and $\sigma$ schemes, and fluid flow directions of each program in spontaneous imbibition experiment and their calculated capillary ($N_{Ca}$), inverse Bond ($N_B^{-1}$), and Weber ($N_{We}$) numbers.

| Sequence name (Flow direction) | Sequential fluids | | $\sigma$ (mN/m) | $N_{Ca}$ | $N_B^{-1}$ | $N_{We}$ |
|---|---|---|---|---|---|---|
| Co-In (co-current) | Fluid I | 0.2 mM | 7.9 | $7.4 \times 10^{-7}$ | 5.1 | $1.3 \times 10^{-14}$ |
| | Fluid II | 0.02 mM | 16.8 | $2.6 \times 10^{-7}$ | 31.3 | $5.9 \times 10^{-15}$ |
| | Fluid III | 0.002 mM | 21.7 | $1.7 \times 10^{-7}$ | 59.8 | $4.6 \times 10^{-15}$ |
| | Fluid IV | 0 mM | 28.0 | $1.1 \times 10^{-7}$ | 116.1 | $3.5 \times 10^{-15}$ |
| Co-De (co-current) | Fluid I | 0 mM | 28.0 | $1.1 \times 10^{-7}$ | 116.1 | $3.5 \times 10^{-15}$ |
| | Fluid II | 0.002 mM | 21.7 | $1.7 \times 10^{-7}$ | 59.8 | $4.6 \times 10^{-15}$ |
| | Fluid III | 0.02 mM | 16.8 | $2.6 \times 10^{-7}$ | 31.3 | $5.9 \times 10^{-15}$ |
| | Fluid IV | 0.2 mM | 7.9 | $7.4 \times 10^{-7}$ | 5.1 | $1.3 \times 10^{-14}$ |
| Ct-In (counter-current) | Fluid I | 0.2 mM | 7.9 | $7.4 \times 10^{-7}$ | *Not applicable due to flow direction.* | $1.3 \times 10^{-14}$ |
| | Fluid II | 0.02 mM | 16.8 | $2.6 \times 10^{-7}$ | | $5.9 \times 10^{-15}$ |
| | Fluid III | 0.002 mM | 21.7 | $1.7 \times 10^{-7}$ | | $4.6 \times 10^{-15}$ |
| | Fluid IV | 0 mM | 28.0 | $1.1 \times 10^{-7}$ | | $3.5 \times 10^{-15}$ |
| Ct-De (counter-current) | Fluid I | 0 mM | 28.0 | $1.1 \times 10^{-7}$ | | $3.5 \times 10^{-15}$ |
| | Fluid II | 0.002 mM | 21.7 | $1.7 \times 10^{-7}$ | | $4.6 \times 10^{-15}$ |
| | Fluid III | 0.02 mM | 16.8 | $2.6 \times 10^{-7}$ | | $5.9 \times 10^{-15}$ |
| | Fluid IV | 0.2 mM | 7.9 | $7.4 \times 10^{-7}$ | | $1.3 \times 10^{-14}$ |

### 2.4.2 Fluid flow directions

Fluid flow directions in spontaneous imbibition were also investigated and coupled with an effect of sequential change in $\sigma$ (Table 2). Two flow directions configured by differentiating boundary conditions of the flow in core sample were considered: co-current and counter-current flows [38]–[42]. Co-current flow in core sample was obtained in all-face-open configuration, where aqueous solution was allowed to displace residing oil to flow in the same direction. Gravitational force also contributes to fluid flow in the all-face-open core sample. On the other hand, counter-current flow only allows the two fluids to move or exchange in opposite directions. This was achieved by sealing top-end and bottom-end of the core sample by epoxy glue (i.e. two-end-open configuration), which prevented the fluid flow in vertical direction while forced the two fluids to flow in and out on the side of the core horizontally. These two flow directions are illustrated in Fig. 2a and 2b, respectively.

As summarized in Table 2, there are four programs for spontaneous imbibition study. The four programs are: (i) co-current flow with sequential decrease in $\sigma$ (Co-De); (ii) co-current flow with sequential increase in $\sigma$ (Co-In); (iii) counter-current flow with sequential decrease in $\sigma$ (Ct-De); and (iv) counter-current flow with sequential increase in $\sigma$ (Ct-In).

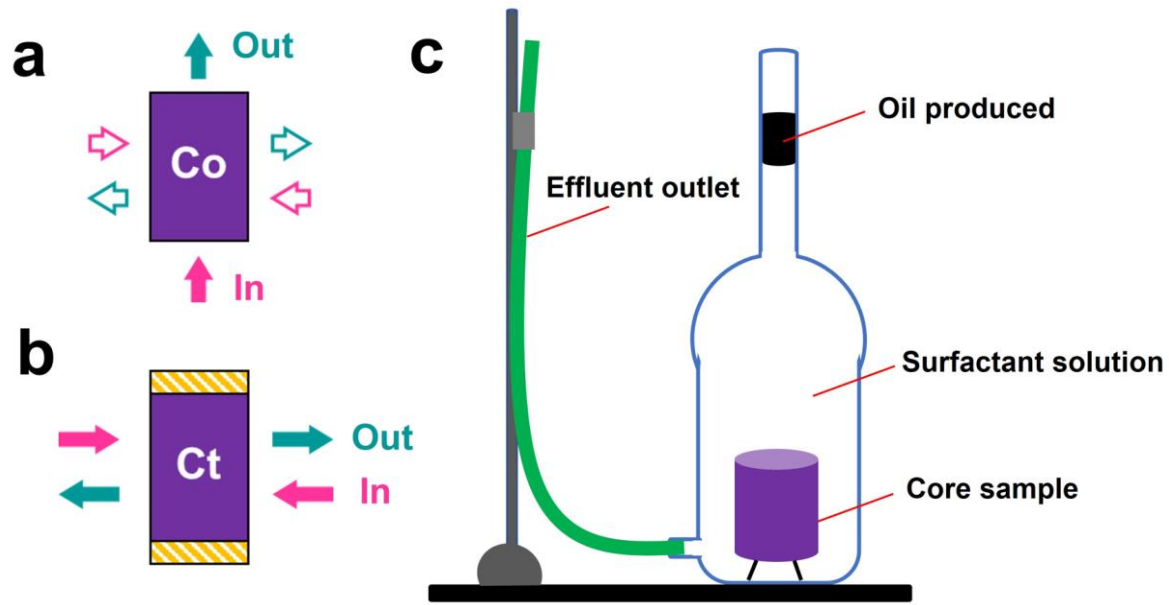

**Figure 2.** Schematic diagram of fluid flow directions: (a) co-current and (b) counter-current flows. Green arrows indicate outer fluid flow and pink arrows indicate inner fluid flow. Hallow arrows indicate weak or negligible horizontal flow in the co-current dominated core sample. Yellow bars indicate no-flow boundary. Spontaneous imbibition in oil-saturated core sample was studied in the modified Amott cell (c).

### 2.4.3 Modified Amott cell and spontaneous imbibition experiment

Since $\sigma$ changing is required, a standard Amott cell was specially modified for the current study to facilitate surfactant fluid changing with effluent outlet connected at bottom part of the cell (Fig. 2c). This allows a present fluid flowing out of the cell through an effluent outlet while being replaced gradually by a new fluid dispensed from a top part of the cell. With outlet and inlet flow rates of the cell remaining equal (at ~150 mL/min), a smooth change of surfactant fluid was conducted with caution while cumulative fluid volume in the cell was kept constant (~700 mL). During fluid changing, no air trapping or bubble partitioning was observed while already-produced oil was stationed steadily.

Prior to experiment, a glass Amott cell was cleaned by rinsed thoroughly with toluene and left dry overnight. Spontaneous imbibition was conducted by placing an oil-saturated core sample in the Amott cell chamber before the cell was steadily filled with imbibing fluid (surfactant solution) up to a scale reading of graduated cylinder. The core sample was thus fully submerged in an imbibing fluid. For co-current flow, a 1-cm-height stainless-wire truss was used to partition a core sample from the Amott cell bottom to allow a vertical flow from bottom-end of the core sample. All imbibition experiments were conducted in an oven at 40 °C. The Amott cell and imbibing fluids were pre-heated in the oven at 40 °C before use. Fluid evaporation from the Amott cell was prevented by sealing the cell and tubing ducts with Parafilm®. Oil produced from a core sample gradually liberated and floated to the top of the aqueous phase at the graduated cylinder scale, where volume of the oil produced was recorded periodically. Until no noticeable change in oil volume was observed, fluid changing was thus performed according to the sequential fluid schemes (Table 2). Oil recovery is reported in percentage of oil volume produced to initial oil saturation.

## 3 Results and Discussion

### 3.1 Reduction in the oil-water interfacial tension and the three-phase contact angle

The interfacial tension between the oil phase and the deionized water (0 mM surfactant) was measured to be 28.0 mN/m. With increasing surfactant concentration (from 0.002 to 0.2 mM Triton X-100) in aqueous phase, the $\sigma$ decreased due to surfactant adsorption at the crude oil-water interface (Fig. 3a). At 0.2 mM surfactant concentration, the lowest $\sigma$ was observed (7.9 mN/m), where the concentration approached the surfactant's CMC [34]. These results demonstrate a benefit of surfactant addition to reduce the $\sigma$.

Surfactant addition was also found to decrease the oil-water-rock contact angle (Fig. 3b). The measured contact angle in the deionized water (0 mM surfactant) was initially 171.8°, indicating a strongly oil-wet nature. Decreased $\theta$ with increasing surfactant concentration (from 0.002 to 0.2 mM) was contributed from a reduction in the $\sigma$, as described by the Young's equation [15]. However, the $\theta$ decrease was not dramatic as the measured $\theta$ in all surfactant concentrations remained in an oil-wet region ($\theta > 120°$), see Fig. 3b.

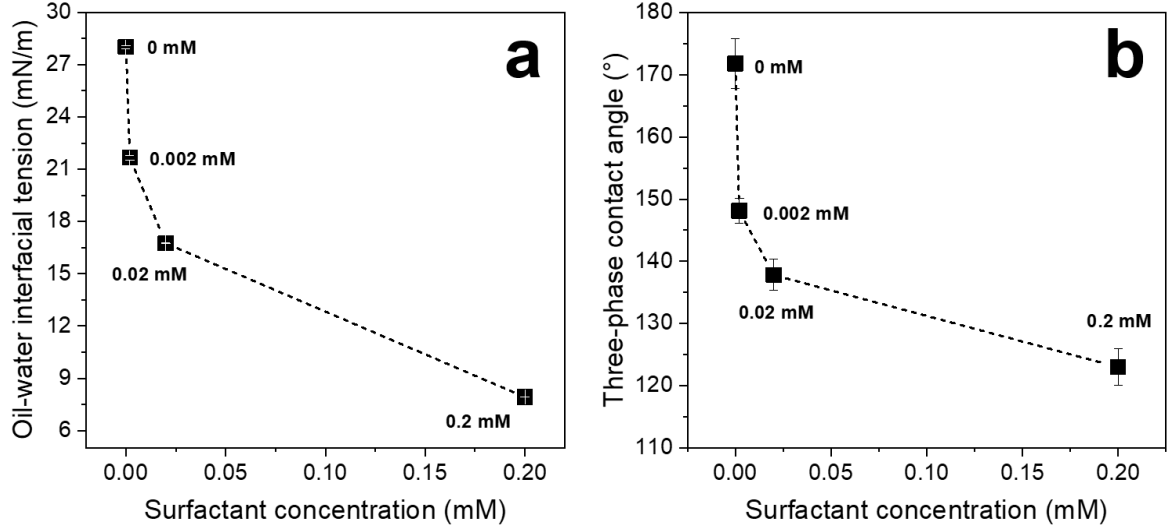

**Figure 3.** The oil-water interfacial tension (a) and the three-phase oil-water-rock contact angle (b) as a function of surfactant concentration. Error bars shown are standard derivation. Lines to guide the eye.

### 3.2 Dominating forces in fluid flow and general observation on spontaneous imbibition

#### 3.2.1 Dimensionless numbers

In the current study, fluid flow in spontaneous imbibition process was primarily under control of two main dominating forces: capillary and gravitational forces [22]. Contribution of each force can be relatively quantified by dimensionless numbers, namely capillary number ($N_{Ca}$) and inverse Bond number ($N_B^{-1}$) [23], [43].

Capillary number ($N_{Ca}$) defines a ratio of viscous to capillary forces, and can be expressed as:

$$N_{Ca} = \frac{\mu V}{\sigma \cos\theta} \qquad (2)$$

where $\mu$ is the oil viscosity, and $V$ the characteristic velocity of the oil.

Inverse Bond number ($N_B^{-1}$) quantifies a ratio of capillary to gravitational forces, as follows:

$$N_B^{-1} = \frac{\sigma \cos\theta}{r \Delta\rho g H} \qquad (3)$$

where $\Delta\rho$ is the density difference between oil and water phases, $g$ the gravity acceleration constant, and $H$ the core length characteristic taken to be the core height.

The two dimensionless numbers in each imbibition and fluid program were calculated from the measured parameters and reported in Table 2, with assuming $V = 1 \times 10^{-6}$ m/s [44]. For all imbibition programs and surfactant concentrations, the distinctively low $N_{Ca}$ ($< 8 \times 10^{-7}$) implies a substantial contribution from capillary force compared to viscous force [45]. Decrease in $\sigma$ at high surfactant concentration only increased $N_{Ca}$ slightly within the same order of magnitude (from $1.1 \times 10^{-7}$ at 0 mM to $7.4 \times 10^{-7}$ at 0.2 mM), thus a viscous force was totally screened in the current study. A negligible viscous contribution was further confirmed with the measured viscosities of the oil and aqueous phases (3.18 and 0.7 mPa·s, respectively) and a low viscosity ratio was obtained (0.22).

Considering a gravitational contribution, only co-current imbibition (Co-De and Co-In) can be analyzed with $N_B^{-1}$ while counter-current flow is assumed to have negligible gravitational contribution due to the rock configuration [46]. The current study of counter-current imbibition (Ct-De and Ct-In) was thus inevitably controlled by only a capillary force (no $N_B^{-1}$ considered).

The calculated $N_B^{-1}$ were greater than 1 in all surfactant concentrations, indicating a relatively weak gravitational contribution over the whole sequence of the co-current imbibition experiments [23], [47]. In the Co-De sequence, decreased $\sigma$ resulted in an obvious decrease in $N_B^{-1}$ (from 116.1 to 5.1), indicating a successive gravitational contribution when aqueous fluids changed from Fluid I to Fluid IV. An opposite $N_B^{-1}$ trend was observed in the Co-In sequence, a contribution from gravitational force was weakened with consecutive increase in $\sigma$ (aqueous fluids changed from Fluid I to Fluid IV). This demonstrates an effect of the oil-water interfacial tension on an interplay between capillary and gravitational forces.

It is noted that the capillary forces in all studied programs acted as a resisting force to oil displacement (i.e. negative capillary) because of an oil-wet nature of the rock-water-oil system remained ($\theta > 90°$, Fig. 3b). Hence, reduction in such a resisting capillary force by reducing $\sigma$ would favor an oil displacement, as previously discussed.

In addition to $N_{Ca}$ and $N_B^{-1}$ numbers, Weber number ($N_{We}$) is introduced to consider an inertia-to-capillary effect [46], [48]:

$$N_{We} = \frac{\rho_w r V^2}{\sigma} \tag{4}$$

where $\rho_w$ is the water density. Table 2 also included the calculated $N_{We}$, while the $N_{We}$ will be later discussed to better elucidate the results in Section 3.3 with an inertia effect.

## 3.2.2 Spontaneous imbibition and its dominating forces

To be accurate, with wettability remained strongly oil-wet (Fig. 3b) oil displacement in the current study was actually a primary drainage which non-wetting phase (water) displaced wetting phase (oil). However, throughout the current work the word imbibition process was used instead to reconcile with a common perception when water is used to displace oil. Results of oil recovery from the four imbibition programs (i.e. co-current and counter-current flows with both sequential changes in $\sigma$) are shown in Fig. 4.

In all programs, one general result was observed with the first fluid. Oil was initially produced at large amount from Fluid I before some incremental oil was subsequently recovered with consecutive replacing fluids (Fluids II – IV). A strong initial imbibition was a result of spontaneous nature of capillarity in porous rock, with a quick water channeling established at a beginning of an imbibition [23], [49], [50]. Smaller initial oil recovery produced in the high $\sigma$ (Fluid I in Co-De and Ct-De programs) was due to stronger resisting capillary ($P_c$ was calculated to be 0.31 MPa), while in the lower $\sigma$ (Fluid I in Co-In and Ct-In programs) larger produced oil was observed with a weaker resisting capillary ($P_c$ was calculated to be 0.08 MPa).

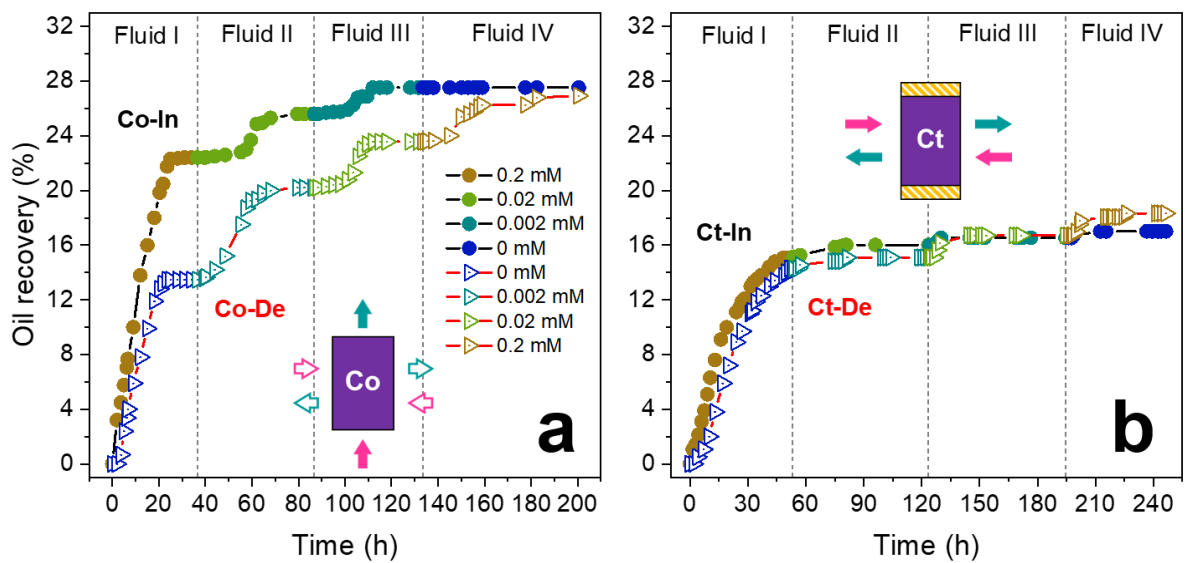

**Figure 4.** Oil recovery from spontaneous imbibition experiments of co-current flows (a) and counter-current flows (b) with sequential increase and sequential decrease in the oil-water interfacial tension.

Considering full imbibition process from Fluids I → IV, oil recoveries from co-current flow (Fig. 4a) were distinctively higher than that of counter-current (Fig. 4b) for both sequential fluid approaches. Higher oil recovery in the co-current flow was a result of additional gravitational contribution on oil displacement, while capillary force remained a principle dominating factor in both flow directions (Table 2). With only capillary contribution, oil recovery in the counter-current program also appeared to be less responsive to the fluid change.

Results of oil recovery were used to calculate an imbibition rate by differentiating the imbibition recovery with time: $\frac{\partial R}{\partial t}$, where $R$ is the oil recovery [23]. As shown in Fig. 5, imbibition rate in each fluid initially increased rapidly before reached a maximum, and then declined to zero indicating that no further oil could be produced. In Fluid I, the imbibition rates in co-current flows (Fig. 5a) were clearly higher than that of counter-current (Fig. 5b). This was due to an additional gravitational force existed only in the co-current flows.

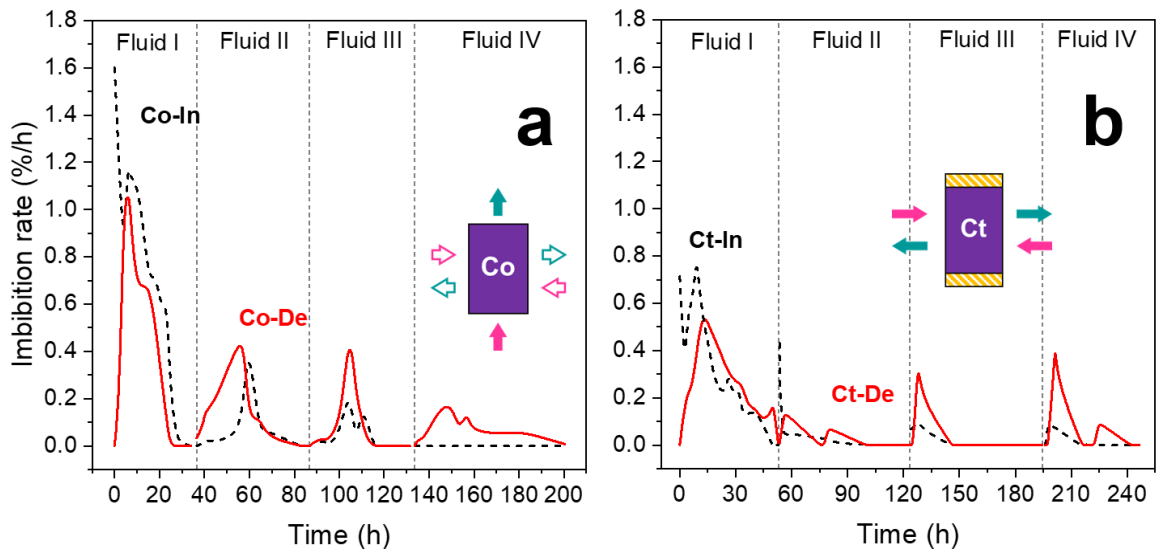

**Figure 5**. Imbibition rate calculated from oil recovery results of co-current flows (a) and counter-current flows (b) with sequential increase (black dashed lines) and sequential decrease (red solid lines) in the oil-water interfacial tension.

With Fluids II – IV, imbibition rates in the sequential decrease approaches (Co-De and Ct-De) were relatively higher than that of the sequential increase (Co-In and Ct-In), especially at maximum imbibition rates. This dynamic behavior agrees with $N_{Ca}$ changes reported in Table 2. Increasing $N_{Ca}$ (by decreasing $\sigma$) gradually faded the resisting capillary contribution, which resulted in a faster oil imbibition rate.

In the following sections, the oil imbibition results are discussed in detail with focusing on two comparisons: (i) effects of single-reduced and sequential-reduced oil-water interfacial tension; and (ii) effects of sequential increase and sequential decrease in the oil-water interfacial tension.

### 3.3 Effect of sequential-reduced oil-water interfacial tension scheme

Effect of sequential reduction in $\sigma$ was examined against a conventional single reduction in $\sigma$, while both schemes obtained the same eventual value of $\sigma$ at 0.2 mM surfactant. Ultimate oil recoveries from the two schemes were compared and shown in Table 3 for both co-current and counter-current flow directions. It is noted that effects of sequential increase in $\sigma$ and single increased $\sigma$ are irrelevant to conventional oilfield development and hence the current study since reducing $\sigma$ by surfactants is normally implemented at the last stage of oil recovery (i.e. no following fluid injection).

**Table 3.** Ultimate oil recovery from single-reduced and sequential-reduced interfacial tension schemes in co-current and counter-current spontaneous imbibition.

| Flow direction | Scheme to reduce the oil-water interfacial tension | Ultimate Recovery (%) |
|---|---|---|
| Co-current | Single-reduced: 0.2 mM | 22.4 |
| | Sequential-reduced: 0 mM → 0.2 mM | 26.9 |
| Counter-current | Single-reduced: 0.2 mM | 15.0 |
| | Sequential-reduced: 0 mM → 0.2 mM | 18.3 |

In co-current flow, the sequential reduction scheme recovered higher ultimate oil (26.9%) than that of the single reduction (22.4%). Although both schemes have experienced the same fluid at final imbibition stage (0.2 mM surfactant solution), dynamic development on $N_{Ca}$ differed. The single $\sigma$ reduction rapidly installed $N_{Ca}$ to the highest (7.4 × 10$^{-7}$), while the sequential reduction scheme allowed $N_{Ca}$ to develop gradually until attained the same highest value (1.1 × 10$^{-7}$ → 7.4 × 10$^{-7}$). A gradual reduction in $N_B^{-1}$ (116.1 → 5.1) was also in the same manner as the $N_{Ca}$ increase. Hence, such a step-wise reduction in $\sigma$ has experimentally appeared to facilitate a greater oil production from spontaneous imbibition.

Dynamic change in $\sigma$ is considered to influence pore-filling events in oil displacement since $\sigma$ is one of the main parameters that controls threshold capillary pressures for those events [51]. Decline in a resisting capillary force favors a piston-like advance in pores and throats as commonly observed in oil-wet porous media [44], [45], [51], [52]. Gradual decline in resisting capillary likely progressed into a gentle invasion-percolation process and thus a connected cooperative pore-filling (Fig. 6a) [52], [53]. This contributed to high liquid-liquid displacing performance. Gentle increase in gravitational force also assisted a pore filling to advance effectively, without any Haines jump or by-passing (Fig. 6b) induced by a rapid change to strong gravity [54]. On the contrary, an immediate and low resisting capillary force with a sudden high gravitational domination rapidly displaced oil phase only in large pores or major well-connected channels. Water invasion would have induced Haines jumps with quick advancing and retracting from narrow throats, and consequently trapped by oil (i.e. Roof snap-off) [45], [55], [56]. Haines jump has normally been observed to occur in early water-flooding process when most pores are occupied by oil only [52]. Consequently, residual oil was left mostly un-displaced due to those unfavorable pore-filling events. These suggests that dynamic change in $\sigma$ (i.e. capillary change rate or 'inertia' of fluid momentum) influenced fluid-fluid displacement behaviors and thus an ultimate oil recovery.

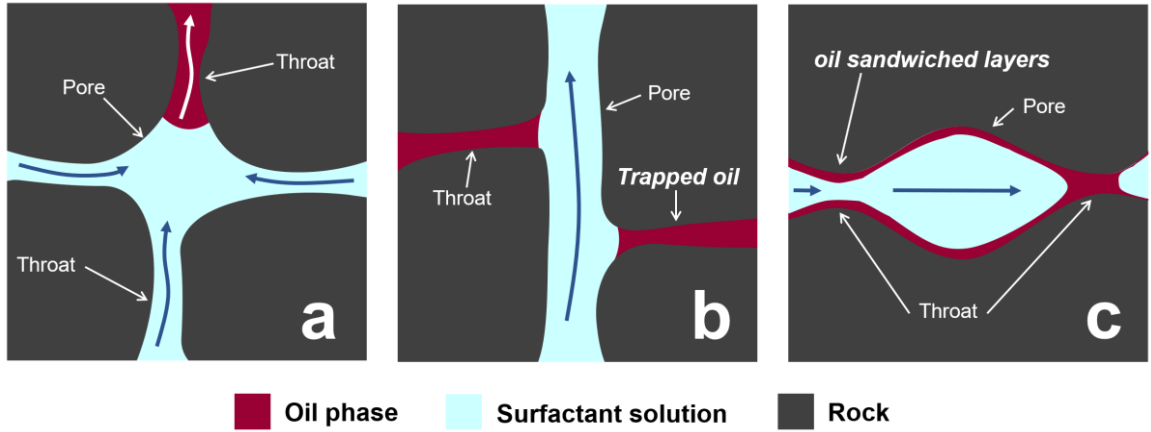

**Figure 6.** Schematic drawings of pore-filling events: cooperative pore-filling (a); Haines jump or by-passing (b); and oil sandwiched layers (c).

To elucidate an inertia effect, the calculated $N_{We}$ (Table 2) were plotted against cumulative oil recovery and shown in Fig. 7. Due to $\sigma$ decrease, an inertia influence was increased accordingly ($N_{We}$ increased). Despite attaining the same $N_{We}$ ($1.3 \times 10^{-14}$ at 0.2 mM surfactant), the cumulative oil recovery from the sequential-reduced interfacial tension scheme was higher than that of the single-reduced scheme (Fig. 7a). This was because the inertia hindrance to oil displacement in the sequential-reduced scheme began at much lower values ($N_{We} < 1.3 \times 10^{-14}$) before later grew with decreasing $\sigma$, allowing some residual oil was continually displaced at less-inertia stage (favors cooperative pore-filling event). In the single-reduced interfacial tension scheme, a high inertia ($N_{We} = 1.3 \times 10^{-14}$) installed readily since the beginning and thus recovered less oil.

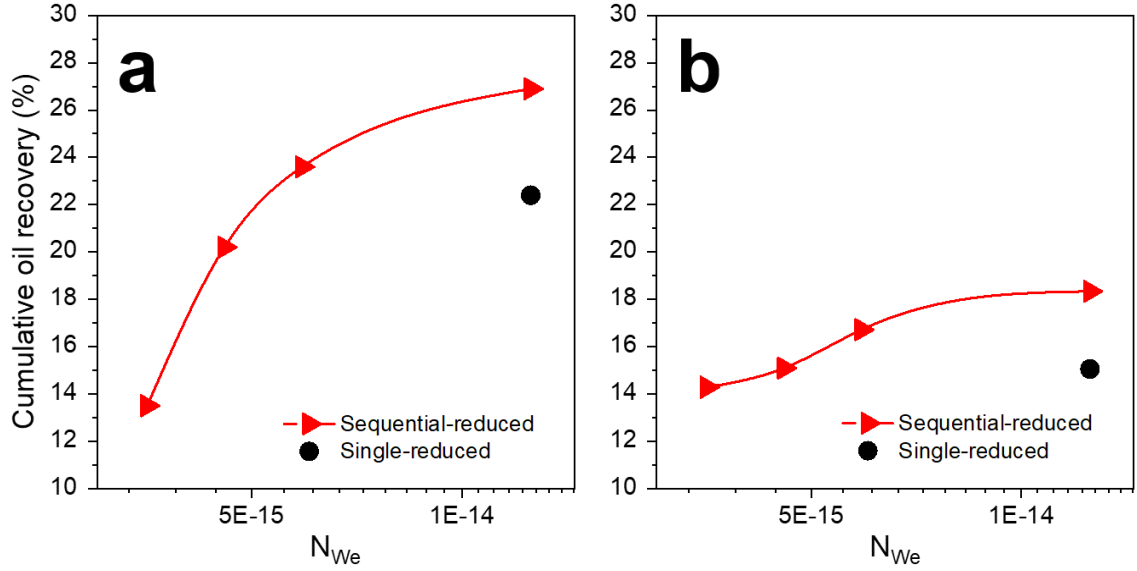

**Figure 7.** Development on cumulative oil recovery with increasing $N_{We}$ in the sequential-reduced interfacial tension schemes (red triangular symbols) compared to that of the single-reduced interfacial tension schemes (black circle symbols) in co-current flows (a) and counter-current flows (b). Lines to guide the eye.

In counter-current flow, the sequential-reduced $\sigma$ scheme also produced higher ultimate oil recovery (18.3%) than the single-reduced $\sigma$ scheme (15.0%) as observed in the co-current direction. With no contribution from gravitational force, oil production in the counter-current direction was relatively low and displaced by only capillary force as mentioned previously.

It is believed that pore-filling events in the counter-current flow also occurred as in the co-current flow, with the same relationship between cumulative oil recovery and $N_{We}$ observed in Fig. 7b. Lower inertia in the sequential-reduced $\sigma$ scheme ($N_{We} < 1.3 \times 10^{-14}$) likely induced a cooperative pore filling displacement, while high inertia hindrance from the single-reduced $\sigma$ ($N_{We} = 1.3 \times 10^{-14}$) contributed to a quick invasion percolation that primarily happened in the main pore channels. Owing to no gravitational force, an inertia effect (the $N_{We}$ change in Fig. 7b) on cumulative oil production were less sensitive to fluid change when compares with the co-current case (in Fig. 7a).

## 3.4 Comparison between sequential increase and sequential decrease in the oil-water interfacial tension

Even though the ultimate oil recoveries were undistinguishable (~27.2%), dynamic oil recoveries from the two sequential approaches differed distinctively in the co-current flow (Fig. 4a). The Co-De program displaced residual oil at substantial amount with every fluid injected, while considerable oil was rapidly produced (~22.4%; 81.4% of the ultimate recovery) by Fluid I in the Co-In program. Imbibition rate plot (Fig. 5a) illustrated such dynamic oil displacement phenomena. Since the two programs experienced opposite route of change in influencing forces (as characterized by $N_{Ca}$ and $N_B^{-1}$ in Table 2), pore-filling events and resulted step-wise oil productions were different even both programs obtained similar oil recovery.

In the Co-In program, early gravitational domination and a weaken resisting capillary force from Fluid I (much reduced $N_B^{-1}$ and great $N_{Ca}$) induced an immediate large oil production as previously discussed. As contributed from a piston-like advance in major paths of large pores [57], such displacement likely not further developed into vicinity pores nearby but rather hindered residual oil in adjacent pores (adjacent pores were not invaded by displacing fluid), and likely then initiated Haines jumps. Roof snap-off (snap-off that happens in drainage process) thus would have also occurred as a consequence of retracting water by the Haines jumps [52], [55]. Water phase was likely blocked (as trapped water ganglions) by wetting phase of oil. Such oil layers that swelled from narrow pore corners to trap water flow were described as 'oil sandwiched layers' by Ekechukwu et al. (Fig. 6c) [45]. Thereafter, the consecutive higher-$\sigma$ fluids (Fluids II – IV) could produce only small amount of incremental oil, see Fig. 4a. With increasing $\sigma$, the trapped oil was also difficult to be displaced by Fluids II – IV because of an increased capillary resisting (Table 2). This investigation demonstrates an advantage of using much decreased-$\sigma$ fluid to displace oil at an early stage, with an inevitable disadvantage of obstructing any secondary fluid displacement.

With sequential lower-$\sigma$ fluids displaced in the Co-De program, resisting capillary force was weakened gradually with continually increased gravitational influence (increased $N_{Ca}$ and decreased $N_B^{-1}$). Such changes likely obtained favorable pore-filling events, a cooperative pore filling in adjacent pores without Roof snap-off, and resulted in incremental oil recovery at each fluid change. Contradicted to the Co-In program, the Co-De program demonstrates an advantage of reducing $\sigma$ to displace oil that begins from higher-$\sigma$ fluid, as analogy to secondary oil production by injecting surfactant solution (lower $\sigma$) after water flooding (higher $\sigma$).

In the counter-current flow (Fig. 4b), same characteristics of oil displacement were observed as in the co-current flow, but less oil recovery (due to no gravitational contribution). Ultimate oil recoveries were about ~17.6% for both Ct-De and Ct-In programs. Absence of gravitational force also influenced dynamic oil productions with no distinctive difference in the two sequential programs.

Figure 8 shows the $N_B^{-1}$-dependent cumulative oil recovery and remaining oil saturation, as analyzed by Schechter et al. [47], from the current results of Co-De and Co-In programs. Induced by reducing $\sigma$, remaining oil saturation is observed to decline as $N_B^{-1}$ decreases [47], [58]. Results from the Co-De program agreed with such reported observation in literature, reflecting a benefit of capillary reduction (with cooperative pore filling displacement). However, a contradictory trend was observed from the Co-In results as increasing $N_B^{-1}$ (by increasing $\sigma$) yielded less remaining oil saturation. This is a consequence of oil sandwiched layers or Roof snap-off event, with less gravitational driven.

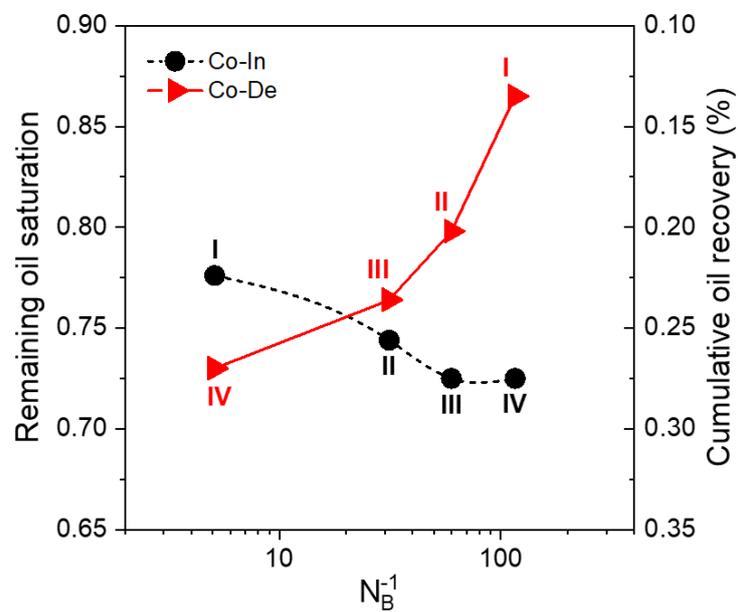

**Figure 8**. Remaining oil saturation and cumulative oil recovery as a function of $N_B^{-1}$ in co-current flows: sequential increase (Co-In, black circular symbols with dashed line) and sequential decrease (Co-De, red triangular symbols with solid line) in the oil-water interfacial tension. Lines to guide the eye.

# 4 Conclusion

As potential attributions from surfactant injection and surfactant retention, continually decreasing and increasing crude-oil water interfacial tension would alter the oil-water-rock capillarity and therefore influence oil displacements in spontaneous imbibition process. The current study has experimentally investigated such changes in the oil-water interfacial tension (by replacing various concentrations of a non-ionic Triton X-100 surfactant) in oil-wet sandstones with both flow directions. Based on the experimental observation and dimensionless numbers analysis, the following conclusions can be drawn:

(1) Governed by a resisting capillary force, continually reduced oil-water interfacial tension was found to produce more incremental oil with higher ultimate recovery, compared to that of a continually increased interfacial tension. Successive oil displacement in the former was thought to be contributed from a cooperative pore filling that facilitated by a weakening capillary resistance (i.e. analogy to surfactant injection), while the latter probably was affected by severe snap-off after a strong piston-like advance displacement that initiated by a very low interfacial tension.

(2) When compared to a conventional one-time reduction in the interfacial tension by using high surfactant concentration, ultimate oil production by a sequential reduction scheme (via replacing fluids consecutively toward higher surfactant concentrations) was observed to be greater. Inertia hinderance to oil displacement induced by capillary change was believed to attribute to such difference. Slowly developing capillary less hindered fluid displacement without immediate piston-like by-passing occurred.

(3) Same characteristics of oil displacement were found in both co-current and counter-current imbibition, even though more oil was produced from co-current imbibition due to assisting gravitational force. Oil recovery results also appeared to more sensitive to fluid changes when gravitational effect included in co-current flow.

**CRediT Author Statement**

**Anupong Sukee**: Methodology, Formal analysis, Investigation, Writing - Original Draft, Visualization. **Tanakon Nunta**: Methodology, Investigation. **Maje Alhaji Haruna**: Investigation, Writing - Review & Editing. **Azim Kalantariasl**: Supervision, Writing - Review & Editing. **Suparit Tangparitkul**: Conceptualization, Methodology, Writing - Review & Editing, Supervision, Project administration, Funding acquisition.22

**Declaration of competing interest**

The authors declare that they have no known competing financial interests or personal relationships that could have appeared to influence the work reported in this paper.

**Acknowledgements**

Financial support for this work was partially contributed from the Murata Science Foundation (20TC04), and Chiang Mai University. The authors would like to thank Northern Petroleum Development Center, Defence Energy Department (Thailand) for providing crude oil sample from Fang oilfields, and also thank Pharit Suya (Chiang Mai University) for his contribution to preliminary experimental work. Undergraduate students in reservoir engineering (256481) class of 2020 (Department of Mining and Petroleum Engineering, Chiang Mai University) are also acknowledged for their enthusiastic discussion.